\documentclass{ifacconf}%
\usepackage{graphicx}
\usepackage{natbib}
\usepackage{mathptmx}
\usepackage{amsmath}
\usepackage{amssymb}
\usepackage{graphicx}
\usepackage{amssymb}
\usepackage{epstopdf}
\usepackage{subfig}
\usepackage{multirow}
\usepackage{amsfonts}%
\providecommand{\U}[1]{\protect\rule{.1in}{.1in}}
\newtheorem{theorem}{Theorem}

\newtheorem{example}{Example}

\begin{document}
\begin{frontmatter}
\title{Robust Fractional order PI Controller for Switching Systems \thanksref{footnoteinfo}}
\thanks[footnoteinfo]{This work was supported by the Spanish Ministry of
Science and Innovation under the project DPI2009-13438-C03.}
\author[First]{S. Hassan HosseinNia},
\author[First]{In\'{e}s Tejado},
\author[First]{Blas M. Vinagre}

\address[First]{Department of Electrical, Electronics and Automation Engineering, School of Industrial Engineering, University of Extremadura, Spain\\ (e-mail: \{hoseinia;itejbal;bvinagre\}@unex.es)}
\begin{abstract}                
This paper is concerned with the robust design for a class of switching systems from a fractional order control (FOC) perspective. In particular, the tuning of the robust fractional order PI (RFPI) controller is proposed based on phase margin and cross over frequency specifications, ensuring the quadratic stability of the system, i.e., the third design specification is given by the stability condition. The quadratic stability is analyzed using a frequency-domain method equivalent to the common Lyapunov one. Two examples are given to show the effectiveness of the proposed controller.
\end{abstract}
\end{frontmatter}

\section{Introduction}

In recent years, the study of switched systems has received a growing
attention. Switched systems are a class of hybrid dynamical systems consisting
of a family of continuous (or discrete) time subsystems and a rule that
orchestrates the switching between them (see e.g.~\cite{Liberzon03}, \cite{Daafouz02}).
The widespread application of such systems is motivated by increasing
performance requirements, and by the fact that high performance control
systems can be realized by switching between relatively simple LTI systems.
However, the potential gain of switched systems is offset by the fact that the
switching action introduces behaviour in the overall system that is not
present in any of the composite subsystems. A survey of main problems in
stability and design of switched systems was reported in \cite{Liberzon99},
\cite{Sun05}, \cite{Sun05b}. There are many methodologies and approaches
developed in the switched systems theory in order to mitigate some problems
encountered in practice (refer to \cite{Dong11} and the references therein for
a current review). However, despite much effort, relatively little work can be
found in the literature from a robustness viewpoint.

Fractional order control (FOC), that is, the generalization to
non-integer-orders of traditional controllers or control schemes, and its
applications, are becoming an important research field since it translates
into more tuning parameters or, in other words, more adjustable time and
frequency responses of the control system, allowing the fulfillment of robust
performances. Robust design of fractional order controllers was investigated
in several papers (see e.g.~\cite{Ladaci09}, \cite{Feliu09}, \cite{Feliu07},
\cite{Ahn07}, \cite{Monje04}). Nevertheless, its application to switching
systems has not been paid much attention, expecting to open a new perspective
in the field.

Given this context, the objective of this paper is to examine stability
issues and robust design of switching systems based on fundamentals of FOC.

The remaining of this paper is organized as follows. Section \ref{BD} provides
some important theorems for quadratic stability of switching systems in the
frequency domain, as well as an introduction to fractional order PI controllers. In Section \ref{problem}, the studied problem is stated.
Section \ref{Robust} addresses the robust design of fractional order PI
controllers for such class of systems. An application example is given in
Section \ref{example} to show the goodness of the proposed controller.
Finally, Section \ref{conclu} draws the conclusions of this paper.

\section{Preliminaries}
\label{BD}
In this section, the quadratic stability of switching systems will be studied. A brief introduction of fractional order PI controllers is also given.

\subsection{Quadratic stability of switching systems}

The quadratic stability of switching systems is studied
using the method introduced in \cite{Karimi_11} in the frequency domain which
is a link between quadratic stability using Lyapunov theory and strictly
positive realness (SPR) properties. Thus, it shows the link between
time-domain conditions and frequency-domain conditions in order to obtain
quadratic stability. 

Consider a switched hybrid system as
\begin{equation}
\dot{x}=Ax,A\in co\left\{  A_{1},...,A_{L}\right\},
\label{Conv}
\end{equation}
where $"co"$ denotes the convex combination and $A_{i}, i=1,...,L$ is the switching subsystem. According to \cite{Pardalos_87},  (\ref{Conv}) can be alternatively written as:
\begin{equation}
\dot{x}=Ax,A=\sum_{i=1}^{L}\lambda_{i}A_{i},\forall\lambda_{i}\geq0,\sum
_{i=1}^{L}\lambda_{i}=1.\label{SWHM}%
\end{equation}

\begin{theorem}
[\cite{Boyd_94}]A system described by (\ref{SWHM}) is quadratically stable if
and only if there exists a matrix $P=P^{T} > 0$, $P \in\mathbb{R}^{n}\times
n$, such that
\[
\label{SWST}A_{i}^{T}P+PA_{i}<0, \forall i=1, ..., L.
\]
\end{theorem}

Denoting
\[
C=%
\begin{bmatrix}
c_{2n-1}, & ..., & c_{1}, & c_{0}%
\end{bmatrix}
,
\]
and
\[
A=%
\begin{bmatrix}
-c_{n-1} & -c_{n-2} & \cdots & -c_{1} & -c_{0}\\
1 & 0 & \cdots & 0 & 0\\
0 & 1 & \cdots & 0 & 0\\
\vdots & \vdots & \ddots & \vdots & \vdots\\
0 & 0 & \cdots & 0 & 1\\
&  &  &  &
\end{bmatrix}
,
\]
a stable polynomial of order $n$ corresponding to the system $\dot{{x}}={A}%
{x}$ is as follows:
\begin{equation}
c(s)=s^{n}+c_{n-1}s^{(n-1)}+\cdots+c_{1}s+c_{0}, \label{TCP}%
\end{equation}

Then the relation between SPRness and the quadratic stability can be stated in
the following theorem.

\begin{theorem}
[\cite{Karimi_11}]Consider $c_{1}(s)$ and $c_{2}(s)$, two stable polynomials
of order $n$, corresponding to the systems $\dot{x}=A_{1}x$ and $\dot{x}%
=A_{2}x$ , respectively, then the following statements are equivalent:

\begin{enumerate}
\item {$\frac{c_{1}(s)}{c_{2}(s)}$ and $\frac{c_{2}(s)}{c_{1}(s)}$ are SPR.}

\item {$\left|  \arg(c_{1}(j\omega)) - \arg(c_{2}(j\omega))\right|  <
\frac{\pi}{2}$ $\forall$ $\omega$.}

\item { $A_{1}$ and $A_{2}$ are quadratically stable, which means that
$\exists P=P^{T}>0\in\mathbb{R}^{n\times n}$ such that $A_{1}^{T}P+PA_{1}<0$,
$A_{2}^{T}P+PA_{2}<0$.}
\end{enumerate}

\label{Freq_stab}
\end{theorem}

\subsection{Fractional order PID controllers}

Fractional calculus (FC) is a more generalized form of calculus. Unlike the integer order calculus, where operations are centered mainly at the integers, FC considers every real number. It owes its origin to a question of whether the meaning of a derivative to an integer order could be extended to still be valid when is not an integer. This question was first raised by L'Hopital on September 30th, 1695. Nowadays, the theoretical and practical interest of fractional order operators is well-established, as well as its applicability to several areas of science and engineering, especially related to automatic control and robotics. 

The widespread use of PID controllers has motivated many researchers to look for better design methods or alternative controllers for many years. As reported in \cite{ChenFOC09}, more than $90$ percent of the control loops are PID-type. The generalization of the PID controller to non-integer order, namely PI$^\lambda$D$^\mu$, was proposed by \cite{Podlubny_99}, whose transfer function, in general form, is given by: 
\begin{equation}
PI^{\lambda}D^{\mu}(s)=k_{p}+\frac{k_{i}}{s^{\lambda}}+k_{d}s^{\mu
},\label{fopid}%
\end{equation}
where $\lambda$ and $\mu$ are the non-integer orders of the integrator and differentiator terms, respectively ($\lambda,\mu\in\mathbb{R}^{+}$). Intuitively, with non-integer order controllers, there are more flexibilities in adjusting the gain and phase characteristics than using integer order ones, i.e., fractional order controllers take advantage of the fractional orders introduced in the control action in order to perform a more effective control. The better performance of this type of controllers in comparison with the integer ones was demonstrated in many works (e.g.~refer to the pioneering study \cite{Podlubny_99}). A review of fractional order controllers can be found in \cite{Monje10}, \cite{Valerio10}, \cite{ChenFOC09}, \cite{Monje08}.

In this paper, we will focus on the fractional order PI controller of the form:
\begin{equation}
K(s)=k_p+\frac{k_i}{s^\alpha},
\label{KS}
\end{equation}
where $\alpha$ is the controller order, $\alpha \in (0,2)$. 

\section{PROBLEM FORMULATION}
\label{problem}

In this section, a more precise problem statement for the class of systems
under consideration is given. 

It is well known a switching system can be potentially destabilized by an
appropriate choice of switching signal, even if the switching is between a
number of Hurwitz-stable closed-loops systems. Even in the case where the
switching is between systems with identical transfer functions, it is
sometimes possible to destabilize the switching system by means of switching
(\cite{Leith03}). Likewise, the concept of robustness with respect to parameter
variations is well defined for LTI systems. However, this issue is somewhat
more difficult to quantify for switched linear systems. In particular,
robustness may be defined with respect to a number of design parameters,
including, not only the parameters of the closed-loop system matrices, but
also with respect to switching signal.

\begin{figure}[ptbh]
\begin{center}
\includegraphics[width=0.5\textwidth]{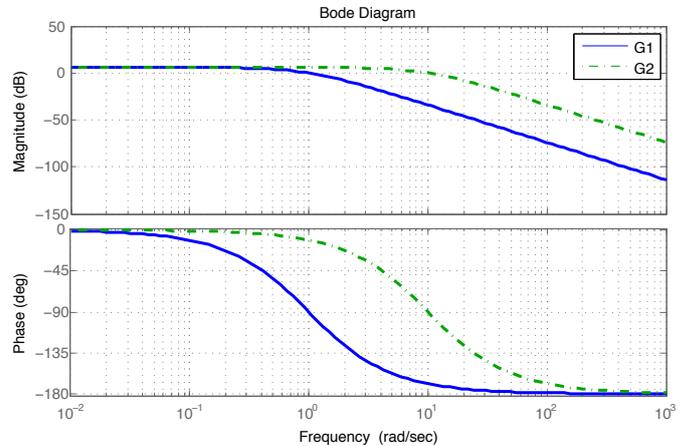}
\end{center}
\caption{Bode diagram of the subsystems $H_1$ and $H_2$}%
\label{FigPIBode}%
\end{figure}

\begin{figure}[ptbh]
\begin{center}
\includegraphics[width=0.5\textwidth]{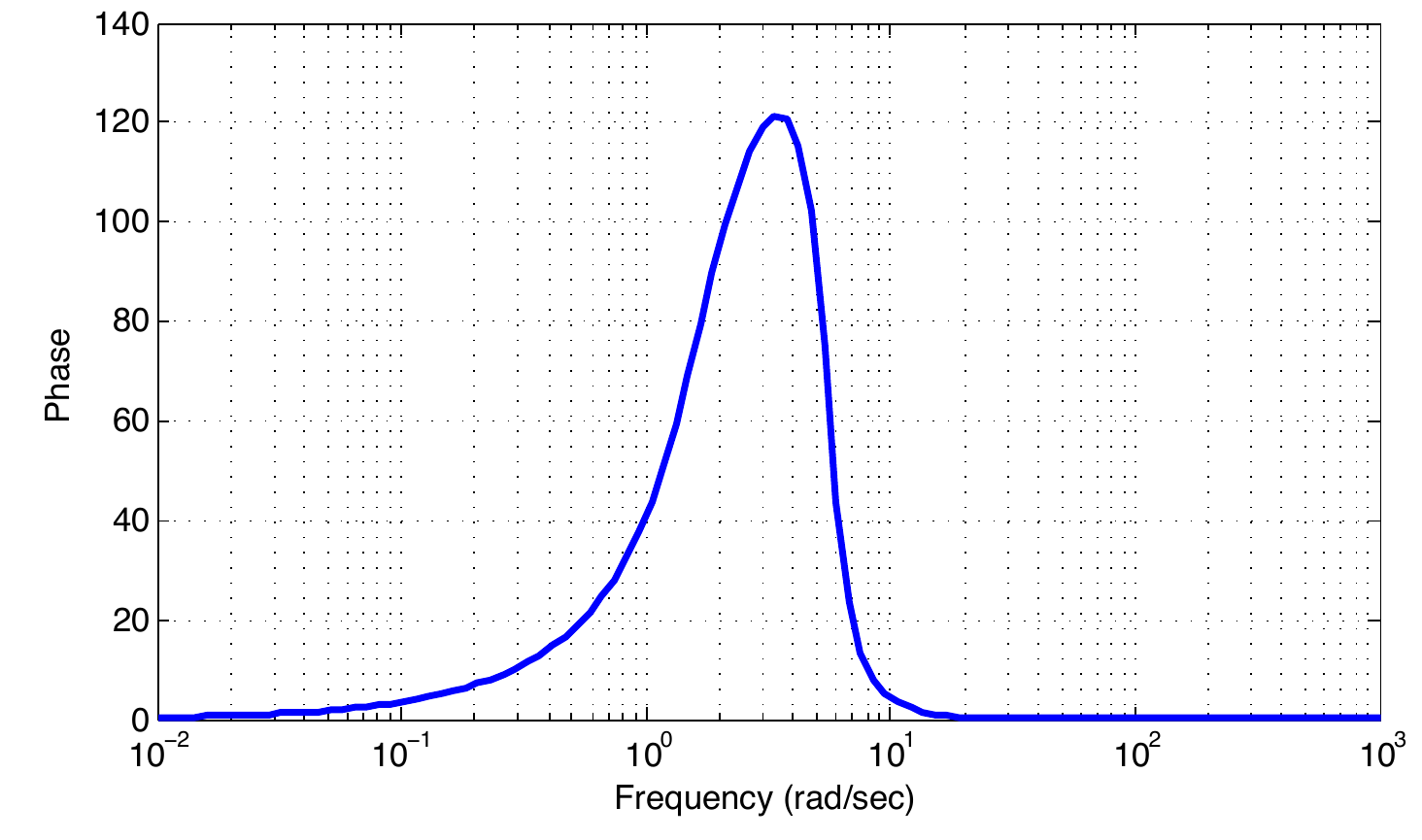}
\end{center}
\caption{Phase difference between the two characteristic polynomials of the
closed-loop subsystems $H_{1}$ and $H_{2}$}%
\label{PIExpCond}%
\end{figure}

Let us illustrate the importance of designing a robust controller for
switching systems by means of a particular example. Consider a switching
system given by the following second order transfer function:
\begin{equation}
H_{i}(s)=\frac{2}{(\tau_{i}s+1)^{2}},\text{ }i=1,2, \label{tf}%
\end{equation}
with $\tau_{1}=1$ and $\tau_{2}=0.1$. One can state than only the time
constant $\tau$ of the system changes. As can be observed in Fig.
\ref{FigPIBode}, both subsystems has the same phase margin of 90 $\deg$.
Applying Theorem \ref{Freq_stab} to the closed-loop system, we have:
\begin{align}
\nonumber \left\vert \arg((3-\omega^{2})+2j\omega)-\arg((3-0.01\omega^{2})+0.2j\omega
)\right\vert  &  \nless\frac{\pi}{2},\\  \label{ConExp}
&  \forall\omega.
\end{align}
Figure~\ref{PIExpCond} depicts (\ref{ConExp}) graphically. Therefore, it is
obvious that the system is not quadratically stable.

\section{Robust Control Design for Switching Systems}
\label{Robust}

\begin{figure}[ptbh]
\begin{center}
\includegraphics[width=0.5\textwidth]{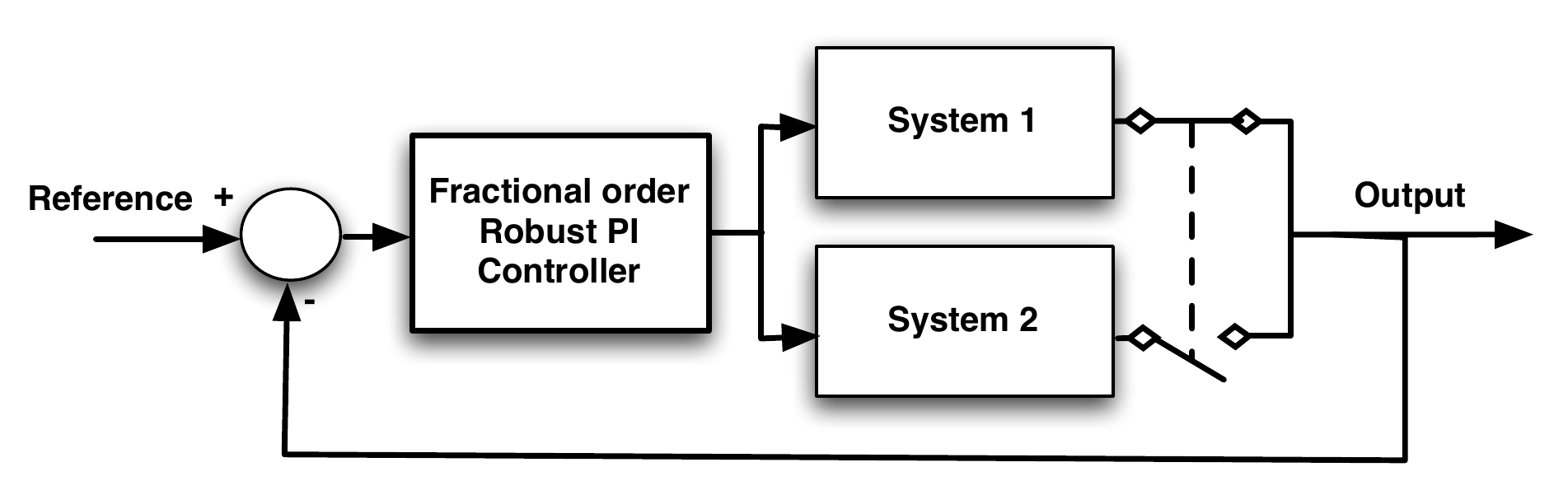}
\end{center}
\caption{Scheme of the controlled system}%
\label{RobustPI}%
\end{figure}

In this section, a method for designing a robust fractional order PI
controller for a class of switching systems is proposed. For ease of the
controller design we consider a system given by (\ref{SWHM}) with $L=2$, but the design procedure can be applied for more than two switching. A scheme of the control approach is shown in Fig.~\ref{RobustPI}. Next, the transfer function of each subsystem will be referred to as $G_{1}(s)$ and $G_{2}(s)$, respectively. Therefore, the aim of this paper is the robust stabilization of a general system $G_{i}(s)$, 
$i=1,2$ by applying a fractional order PI controller of the form (\ref{KS}) --will be referred as to RFPI in the text--, i.e., cancelling the effect of switching on the system stability. 

Let assume that the phase margin and gain crossover frequency are denoted as $\phi_{m}$ and $\omega_{cp}$, respectively. Thus, the three specification to be fulfilled are:

\begin{enumerate}
\item Phase margin specification:
\begin{equation}
Arg(K(j\omega_{cp_{i}})G_{i}(j\omega_{cp_{i}}))+\pi>\phi_{m},\label{SPe1}
\end{equation}
\item Gain crossover frequency specification:
\begin{equation}
\left|  K(j\omega_{cp_{i}})G_{i}(j\omega_{cp_{i}}) \right|  _{dB}=0dB,
i=1,2,\cdots, L. \label{SPe2}%
\end{equation}
\item Quadratic stability specification:
\begin{equation}
\left\vert \arg(c_{1}(j\omega))-\arg(c_{2}(j\omega))\right\vert <\frac{\pi}%
{2},\forall\omega
\label{cri3}
\end{equation}
where $c_{1}$ and $c_{2}$ are characteristic polynomials of each closed-loop system. In order to apply this condition to design the fractional order PI controller,  CRONE approximation of fractional order integral is used. One can use any other approximation for fractional order integral.
\end{enumerate}

To determine the controller parameters, the set of nonlinear equations (\ref{SPe1})-(\ref{cri3}) has to be solved. To do so, the optimization toolbox of
Matlab can be used to reach out the better solution with the minimum error. More precisely, the function FMINCON is able to find the
constrained minimum of a function of several variables. It solves problems of
the form $\min_{x}f(x)$ subject to: $C(x) \leq 0$, $C_{eq}(x)=0$, $x_m
\leq x\leq x_M$, where $f(s)$ is the function to minimize; $C(x)$ and
$C_{eq} (x)$ represent the nonlinear inequalities and equalities, respectively
(non-linear constraints); $x$ is the minimum we are looking for; and $x_m$ and
$x_M$ define a set of lower and upper bounds on the design variables, $x$.

\section{Examples}
\label{example}

This section gives two examples of application of the proposed method for designing robust fractional order PI controllers for switching systems. Specifically, two cases will be considered next: the velocity control of a car, described by a first order model, and the control of a second order system. 

\begin{example}
In \cite{HosseinNia12} and \cite{HosseinNia_11}, the authors proposed a hybrid
model of the car taking into account its different dynamics when accelerating and braking. In this respect, its throttle and brake dynamics can be given, respectively, as follows:
\begin{equation}
G_{1}(s)\simeq\frac{4.39}{s+0.1746},\label{throttle}%
\end{equation}
and
\begin{equation}
G_{2}(s)\simeq\frac{4.45}{s+0.445}.\label{brake_eq}%
\end{equation}

For the comfort of car's occupants, phase margin and crossover frequency has to be chosen around $80$ $\deg$ and $0.8$ rad/s, respectively, in order to obtain
a smooth closed-loop response, i.e., with a value of overshoot $M_{p}$ close to $0$.
Solving the set of equations (\ref{SPe1})-(\ref{cri3}) for the previous specifications, the PI controller is given by:
\begin{equation}
K(s)=0.15+\frac{0.07}{s^{0.71}}.\label{ForPIt}%
\end{equation}
Figure~\ref{Car_Stab_FRPI} shows the frequency response of
the controlled system. As can be seen, the design specifications are fulfilled for both subsystems --the phase margin obtained after optimization problem is even higher
than $80$ deg. 

The phase difference between the two characteristic polynomials of the closed-loop controlled subsystems is shown in Fig.~\ref{Car_FRPI_cond}. It is observed that the maximum phase difference is $27.35$ deg --less than $90$ deg--, so the controlled system is quadratically stable. 

\begin{figure}[ptbh]
\begin{center}
\includegraphics[width=0.5\textwidth]{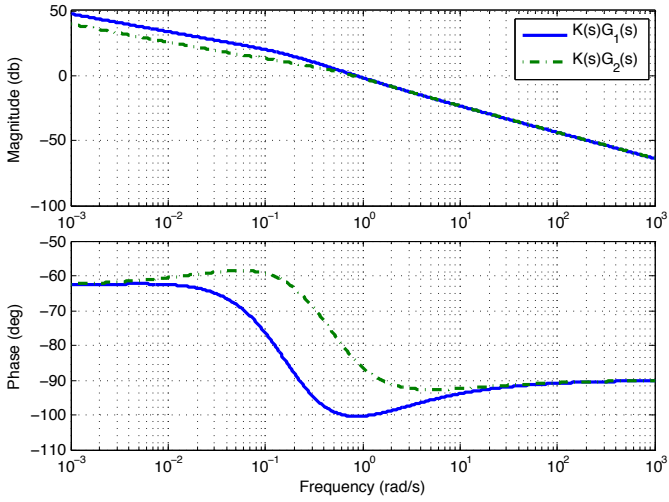}
\end{center}
\caption{Bode plots of the controlled system when applying RFPI.}%
\label{Car_Stab_FRPI}%
\end{figure}
\begin{figure}[ptbh]
\begin{center}
\includegraphics[width=0.5\textwidth]{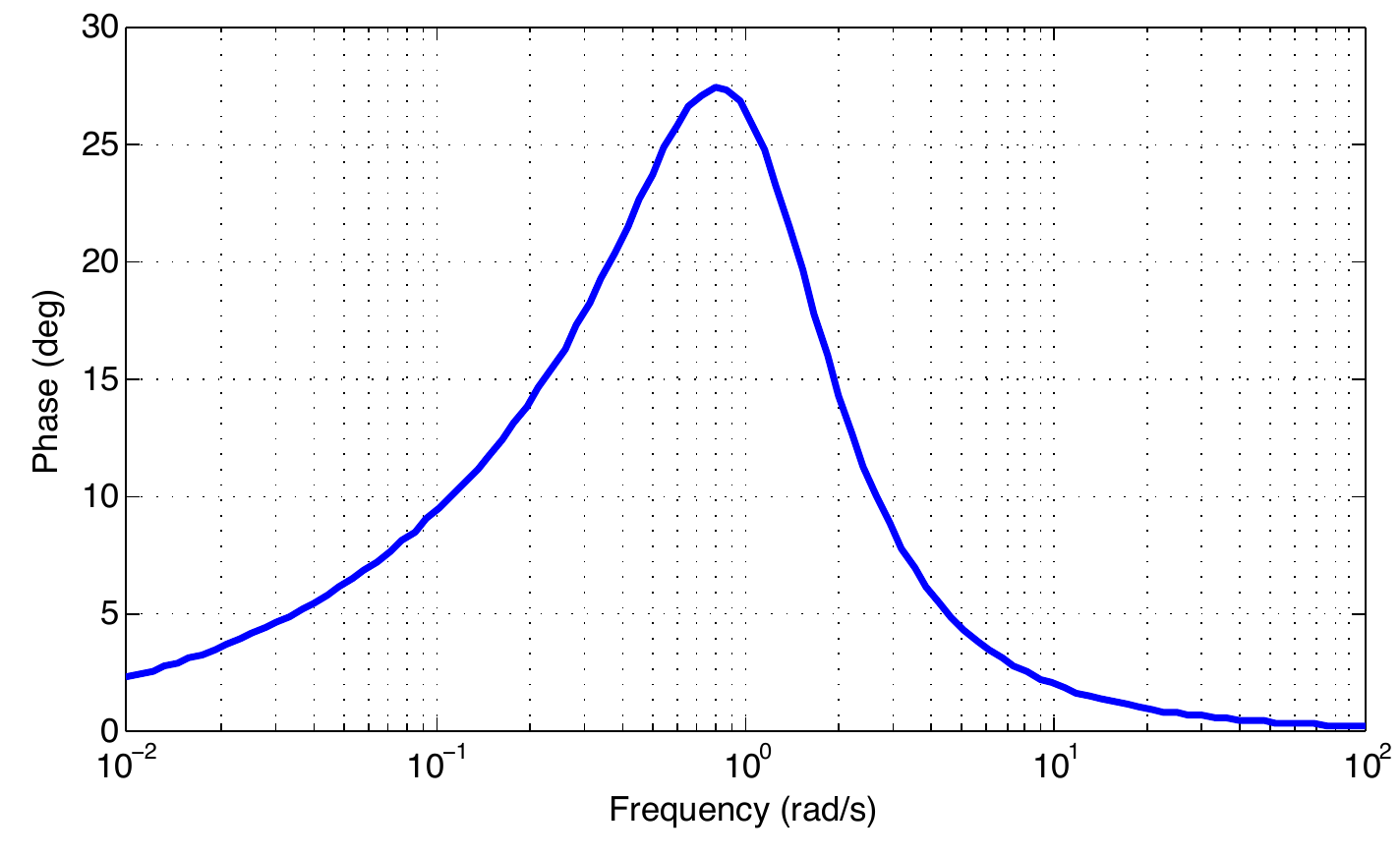}
\end{center}
\caption{Phase difference between the two characteristic polynomials of the
closed-loop system when applying RFPI.}%
\label{Car_FRPI_cond}%
\end{figure}

For comparison purposes, a three-parameter controller, specifically, a traditional PID controller of the form
\begin{equation}
K(s)=k_p+\frac{k_i}{s}+k_ds,
\label{PIDt}%
\end{equation}
is designed for the same specifications (\ref{SPe1})-(\ref{cri3}). Applying the optimization method, the PID parameters are: $k_p=0.1$, $k_i=0.11$ and $k_d=0.223$. Figures~\ref{Exp1PIDbode} and \ref{Exp1PIDSW} show the bode diagram of each closed-loop subsystem when applying the PID and phase difference between the two characteristic polynomials of the closed-loop controlled subsystems, respectively. The phase margin for both subsystems and the maximum of the phase difference are about $80$ and $10.57$ deg, respectively, so the design specifications are fulfilled. Comparing the Bode plots of both controlled system when applying RFPI and PID, i.e., Fig.~(\ref{Car_Stab_FRPI}) and (\ref{Exp1PIDbode}), one can observe that the system controlled by PID has constant magnitude for high frequency, which may cause the system sensitive to high frequency noises and, consequently, instability.

\begin{figure}[ptbh]
\begin{center}
\includegraphics[width=0.5\textwidth]{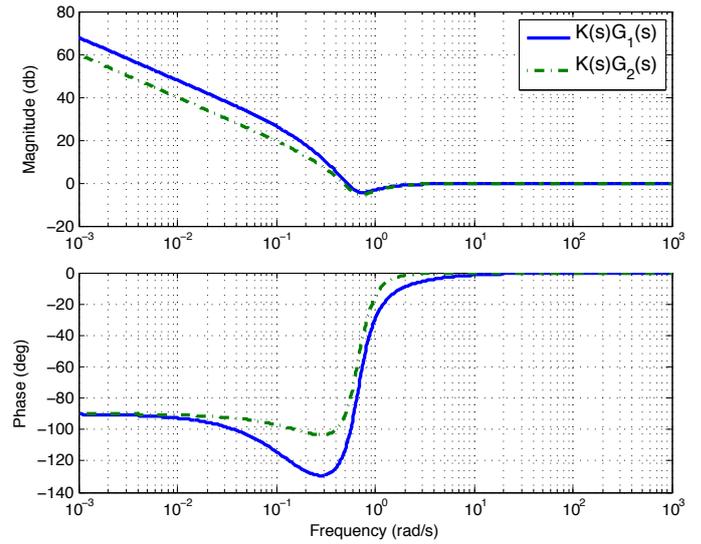}
\end{center}
\caption{Bode plot of the controlled system when applying PID.}%
\label{Exp1PIDbode}%
\end{figure}

\begin{figure}[ptbh]
\begin{center}
\includegraphics[width=0.5\textwidth]{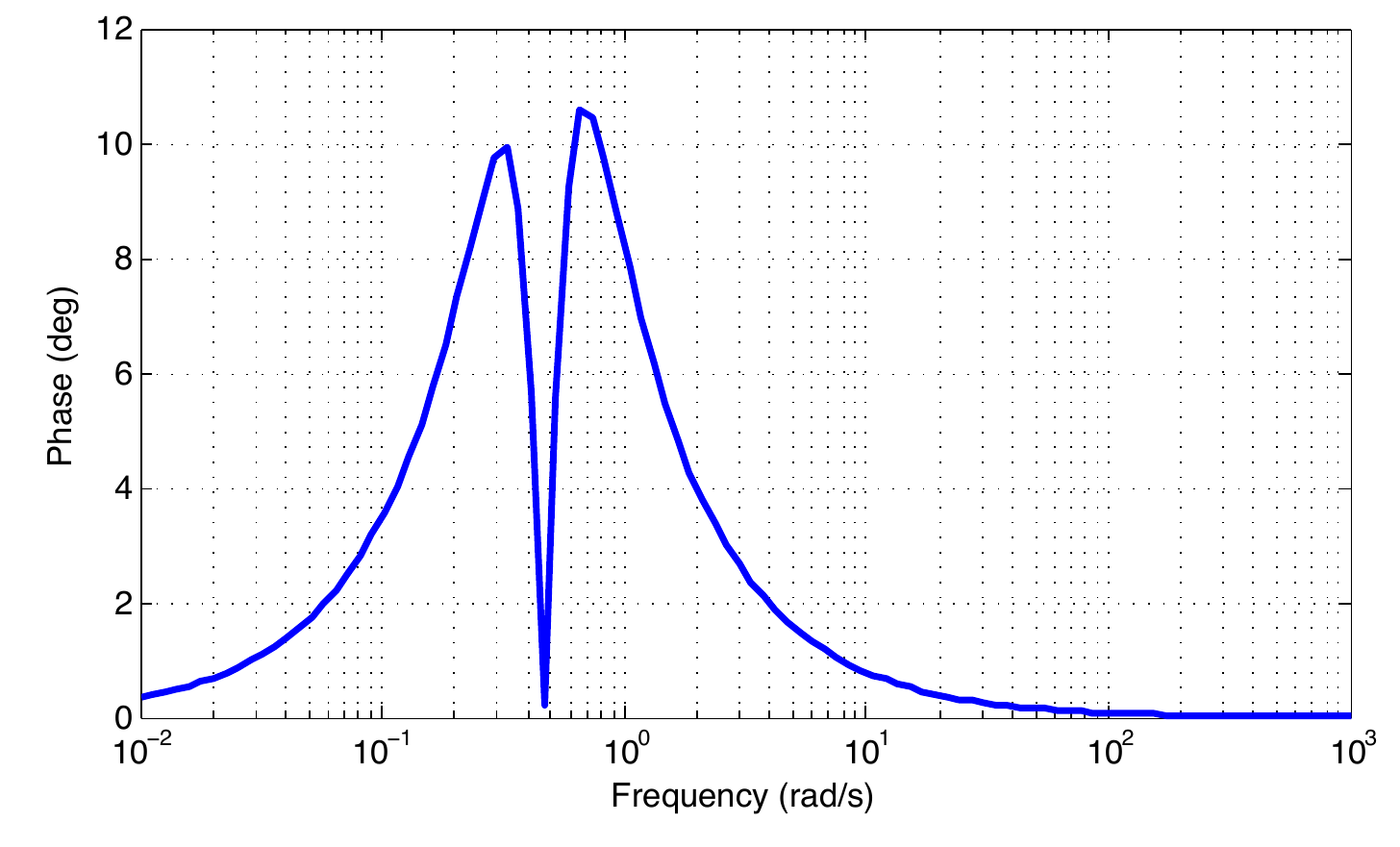}
\end{center}
\caption{Phase difference between the two characteristic polynomials of the
closed-loop system when applying PID.}%
\label{Exp1PIDSW}%
\end{figure}

To show the system performance in time domain, a maneuver which simulates the car acceleration to $20$ km/h and, after that, the braking to $0$ km/h --stop completely-- is depicted in Fig.~\ref{Car_Timedomain} in (a) and the normalized control action when using the RFPI in (b). In addition, for arbitrary switching, including a comparison of its velocity for the RFPI and PID cases is depicted in Fig.~\ref{Exp1comp}. As observed in this figure, the car has an adequate performance for both the throttle and the brake actions when applying the RFPI controller (dashdotted black line), achieving the reference velocity in a suitable time and without overshoot in both cases. On the contrary, the response when using the PID (dashed red line) is significantly poor --has considerable overshoot--. Likewise, as shown in Fig.~\ref{Car_Timedomain} (b), throttle and brake actions for the RFPI case are low, which will be referred to soft actions. As a result, it can be said that the occupants' comfort is guaranteed when applying the proposed controller. It is important to remark that the control action when applying the PID was omitted due to the bad behavior of the system obtained by this controller. It should be mentioned that in the real application the car will not have negative velocity, and in this case when the velocity reach zero the car will stop.

\begin{figure}[ptbh]
\begin{center}
\includegraphics[width=0.5\textwidth]{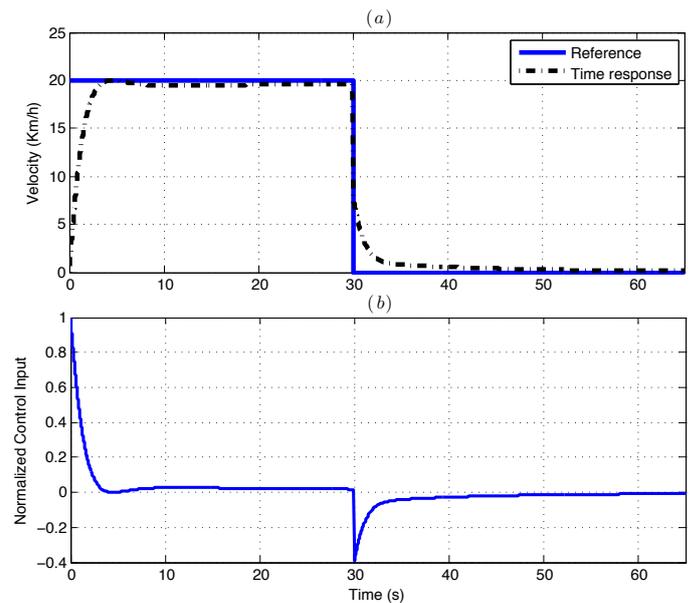}
\end{center}
\caption{Car performance when applying RFPI: (a) Velocity (b) Normalized control action.}%
\label{Car_Timedomain}%
\end{figure}

\begin{figure}[ptbh]
\begin{center}
\includegraphics[width=0.5\textwidth]{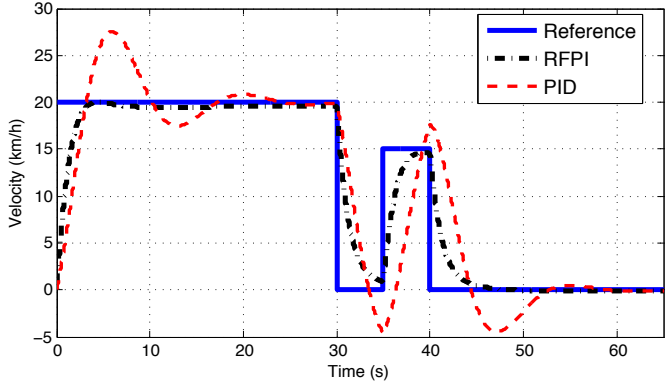}
\end{center}
\caption{Time response of the both controlled system with PID and RFPI during random switching.}%
\label{Exp1comp}%
\end{figure}

\end{example}

\begin{example}
This example considers a second order switched system given by:
\begin{eqnarray}
G_{1}(s)=\frac{2}{s^2+0.6s+0.1},\\ \label{sys1} 
G_{2}(s)=\frac{0.5}{s^2+0.3s+0.023}.\label{sys2}
\end{eqnarray}
The design specifications for the RFPI controller are as follows: $\phi_m>75$ deg, $\omega_{cp_1}\simeq0.6$ rad/s and $\omega_{cp_2}\simeq0.15$ rad/s. After the optimization process, the RFPI obtained is: 
\begin{eqnarray}
K(s)=0.45+\frac{0.05}{s^{1.38}}.
\end{eqnarray}
The Bode plots of the controlled system is depicted in Fig.~\ref{Exp2bode}. As observed, the phase margins are $\phi_{m_1}\simeq 80>75$ deg and $\phi_{m_2}\simeq97>75$ deg, which fulfill the design specifications. The phase difference between the two characteristic polynomials of the closed-loop controlled system is plotted in Fig.~\ref{Exp2_SW}. It can be seen that the maximum value of phase difference is $47.47$ deg, then the system is quadratically stable. 

\begin{figure}[ptbh]
\begin{center}
\includegraphics[width=0.5\textwidth]{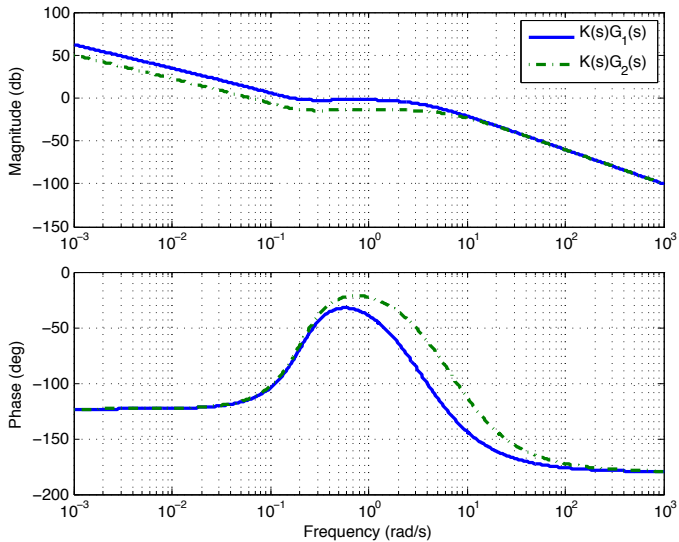}
\end{center}
\caption{Bode diagram of each controlled subsystems.}%
\label{Exp2bode}%
\end{figure}

\begin{figure}[ptbh]
\begin{center}
\includegraphics[width=0.5\textwidth]{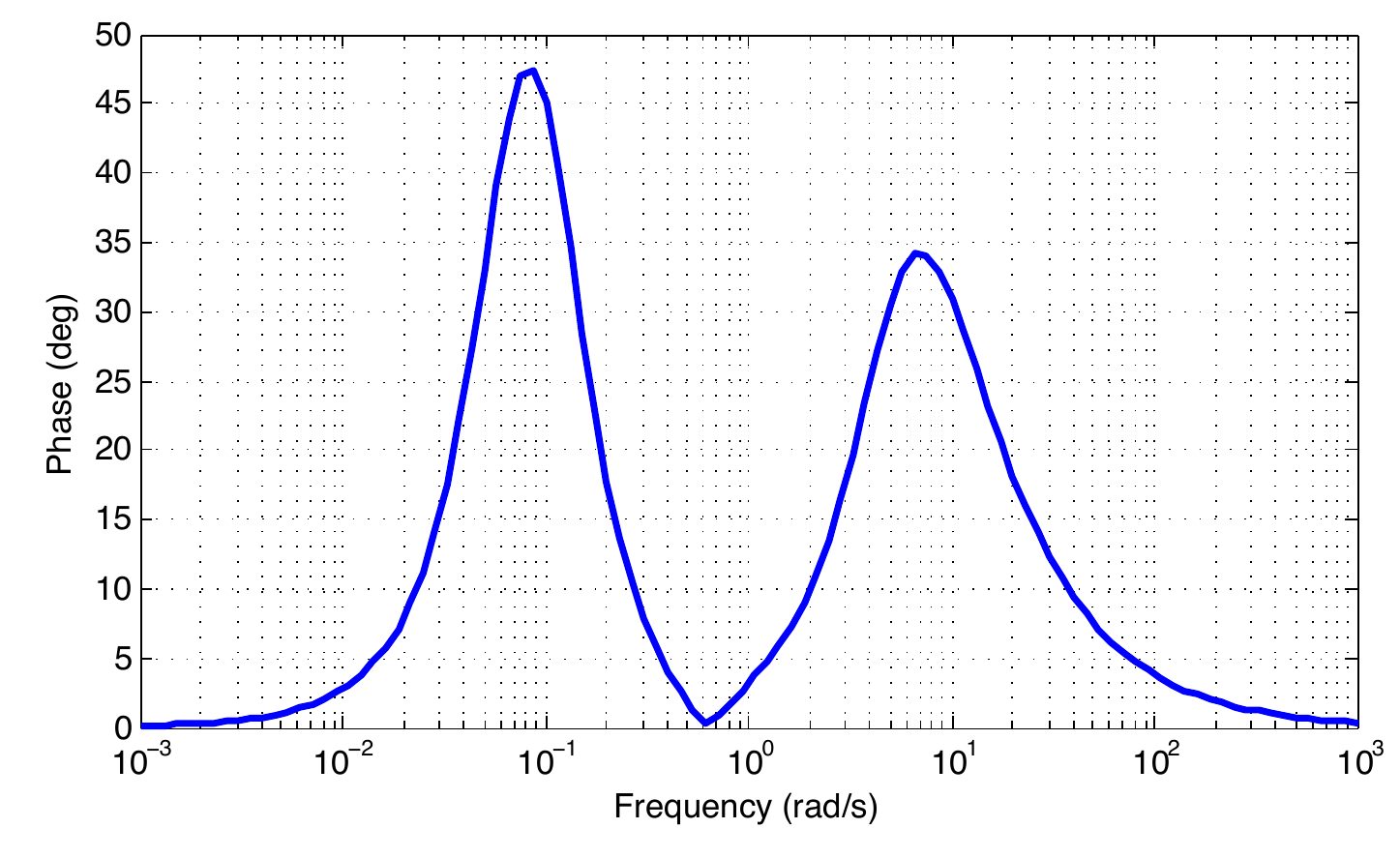}
\end{center}
\caption{Phase difference between the two characteristic polynomials of the
closed-loop controlled subsystems.}%
\label{Exp2_SW}%
\end{figure}

In order to demonstrate the goodness of the designed RFPI controller against switching, two situations will be simulated next: arbitrary and fast arbitrary switching. The step responses are shown in Fig.~\ref{Exp2_TRn} and \ref{Exp2_TR}, respectively. Switching signal is set to $1$ or $0$, which indicates the subsystem $G_{1}$ or subsystem $G_{2}$ is activated, respectively. One can see that the system responses go through an adaptation when the system switches, but achieving the reference and keeping stable for both switching situations.

\begin{figure}[ptbh]
\begin{center}
\includegraphics[width=0.5\textwidth]{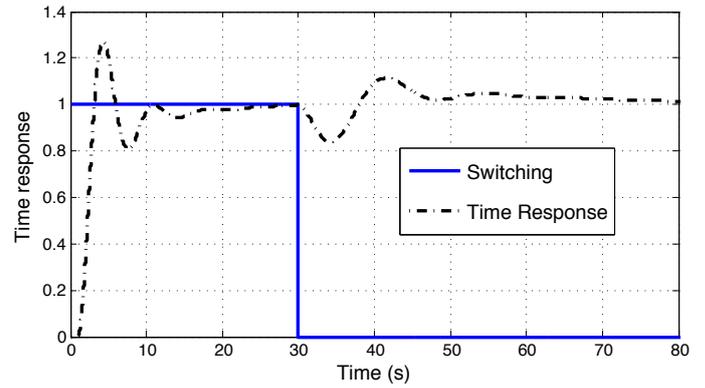}
\end{center}
\caption{Step response of the system for arbitrary switching.}%
\label{Exp2_TRn}%
\end{figure}

\begin{figure}[ptbh]
\begin{center}
\includegraphics[width=0.5\textwidth]{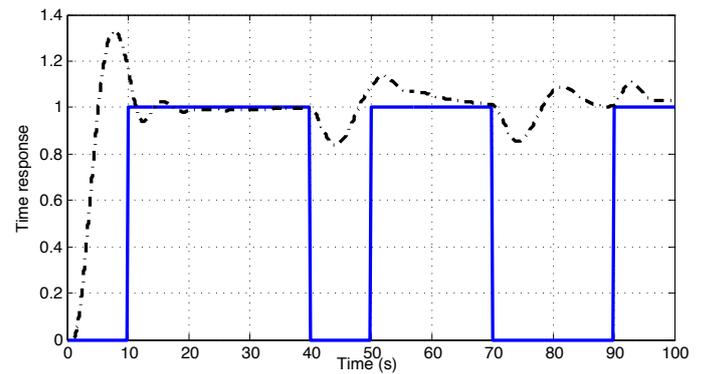}
\end{center}
\caption{Step response of the system for fast switching.}%
\label{Exp2_TR}%
\end{figure}
\end{example}

\section{Conclusion}
\label{conclu}

This paper addresses the design of a robust fractional order PI (RFPI) controller for
switching systems in frequency domain. To this respect, RFPI is tuned based on phase margin and cross over frequency specifications, ensuring the quadratic stability of the system, i.e., the third specification is given by the stability condition. Two examples for switching systems with $2$ subsystems are given to show the effectiveness of the proposed controller. In particular, the speed control of the car is considered, in which the car has different dynamics for accelerating and braking actions. The other example refers to a second order switched system. In both examples, the goodness of the RFPI controller is demonstrated against arbitrary switching. Especially, it is worth mentioning the relevant system performance when applying RFPI in comparison with the obtained by a traditional PID controller, which was also designed to fulfill similar frequency-domain specifications.

Our future efforts will focus on the generalization of the RFPI design for switching systems with $m$ subsystems and its experimental validation.

\bibliographystyle{plain}
\bibliography{ifacconf}








\end{document}